\documentclass[aps,showpacs,floats]{revtex4}
\usepackage{amssymb}
\usepackage{graphicx}
\usepackage{bm}

\begin{document}

\bibliographystyle{apsrev}

\def\v#1{{\bf #1}}
\newcommand{\vq}{{\bf q}}
\newcommand{\vR}{{\bf R}}
\newcommand{\tv}{\tilde{v}}
\newcommand{\tom}{\tilde{\omega}}
\newcommand{\tg}{\tilde{g}}
\newcommand{\tR}{\tilde{R}}

\newcommand{\la}{\langle}
\newcommand{\ra}{\rangle}
\newcommand{\lc}{\lowercase}

\newcommand{\no}{\nonumber}
\newcommand{\be}{\begin{equation}}
\newcommand{\ee}{\end{equation}}
\newcommand{\bea}{\begin{eqnarray}}
\newcommand{\eea}{\end{eqnarray}}

\newcommand{\tphi}{\tilde{\varphi}}
\newcommand{\tthe}{\tilde{\vartheta}}
\newcommand{\trho}{\tilde{\rho}}
\newcommand{\tE}{\tilde{E}}
\newcommand{\hv}{\hat{v}}
\newcommand{\ho}{\hat{\omega}}
\newcommand{\ham}{\hat{m}}

\newcommand{\Q}{(0,0,$\frac{\pi}{c}$)}
\newcommand{\Qh}{(0,0,$\frac{\pi}{2c})$}

\newcommand{\TGG}{T\lc{b}$_3$G\lc{a}$_5$O$_{12}$}

\title{Paramagnetic Acoustic Faraday Rotation in \TGG}

\author{ Peter Thalmeier}
\affiliation{Max Planck Institute for Chemical Physics of Solids, 01187 Dresden, Germany}

\begin{abstract}
The acoustic Faraday rotation in the 4f paramagnet Tb$_3$Ga$_5$O$_{12}$ has recently been observed by Sytcheva et al (arXiv:1006.0141). As in earlier examples the rotation angle per unit length of transverse acoustic modes was found to depend linearly on sound frequency. Existing theories for this effect consistently require that it should vary with the square of the frequency. In the present work a solution for this long-standing problem is provided. We propose a model based on magnetoelastic interactions with 4f quadrupole moments that includes both acoustic and optical phonons. The symmetry allows a direct and induced coupling between the latter. This leads to an indirect acoustic Faraday rotation via the field induced splitting of doubly degenerate optical phonon modes. It varies {\it linearly} with frequency in accordance with experiment and dominates the rotation angle. It also explains the observed resonance of the rotation angle in the field range between 17-20 T. The mechanism is of general validity for non-Bravais lattices and applies to previous examples of the acoustic Faraday effect.
\end{abstract}


\maketitle


\section{Introduction}
\label{sect:introduction}

In crystals with a fourfold uniaxial or higher symmetry transverse acoustic phonons which propagate along axis direction are twofold degenerate with perpendicular polarizations of the displacement vector {\bf u}. Equivalently they may be described by left (L)- and right (R) handed circularly polarized phonons which belong to complex conjugate representations of the  group $G_{\v q}$ of the phonon wave vector \v q which is aligned with an axis. The degeneracy of these complex conjugate modes is ensured by time reversal invariance. If the latter is broken by application of a magnetic field the degeneracy may be lifted. Depending on field direction this leads to the Faraday rotation for \v H $\parallel$ \v q  or Cotton-Mouton effect for \v H $\perp$ \v q in the long-wave length limit. Only the former will be considered here. This means that for fixed sound wave frequency $\omega_a$ the wave numbers $q_L$ and $q_R$ of circular polarised modes will be different. In an ultrasonic experiment a linearly polarised transverse mode is generated which may be described as a superposition of circular modes with equal amplitudes and phase shift $\pi/2$. Since the latter have different wave numbers this means that the polarisation direction of the linearly polarized mode will be continuously rotated as it propagates along the crystal axis. The displacement field of the acoustic wave is then given by
\be
\v u(z,t) = A(\hat{\v x}\cos\phi+\hat{\v y}\sin\phi)\exp(i\omega_a t-iqz)
\label{eq:wave}
\ee
Where A is the amplitude, z the propagation (axis) and x,y the polarization directions. Furthermore average wave number $q$ and  Faraday rotation angle $\phi$ are given by
\be
q=\frac{1}{2}(q_L+q_R)\qquad \phi=\frac{1}{2}(q_L-q_R)z \equiv \phi'z
\ee
Here $\phi'(H)$ is the Faraday rotation angle per unit length.\\

The splitting between $q_L$ and $q_R$ wave numbers due to time reversal symmetry breaking will only be effective when the sound waves can couple strongly to magnetic degrees of freedom. Therefore the acoustic Faraday effect was first observed in magnetically ordered 3d compounds  such as yttrium-iron garnet (YIG) \cite{Matthews62,Guermeur67} and in paramagnets with magnetic impurities \cite{Guermeur68,Kochelaev62,Tucker80}. In the former it is due to magnetoelastic coupling caused by the strain dependence of the anisotropy energy of ordered moments \cite{Kittel58}. It has also been proposed \cite{Tucker73} for S=1/2 paramagnets and for paramagnetic 4f compounds with general  CEF split level scheme \cite{Thalmeier78}.  In the latter case the Faraday effect is due to the magnetoelastic coupling to (2J+1) crystalline electric field (CEF) split  4f states where J is the total angular momentum. This mechanism was first found in the paramagnetic phase of  CeAl$_2$ \cite{Luethi79}. In a magnetic field \v H $\parallel$ \v q  the magnetoelastic coupling leads to nondiagonal quadrupolar susceptibilities of the 4f CEF states resulting in a Faraday rotation of the sound wave polarisation. Both mechanisms lead to a Faraday rotation angle $\phi'$ which increases quadratically with frequency $\omega_a$. However experimentally in CeAl$_2$ the rotation angle turned out to increase linearly with sound frequency. This discrepancy has been unresolved sofar. Recent Faraday rotation experiments on paramagnetic \TGG~ or terbium-gallium garnet (TGG) (Ref. \onlinecite{Sytcheva10}) gave the same linear frequency dependence. Therefore it is clear that the existing theories need to be extended to explain these experimental findings. We note that the Faraday rotation angle in the magnetically ordered systems like YIG was also predicted  to vary with the square of the sound frequency \cite{Matthews62,Guermeur68}. However this has never been tested experimentally because only a single frequency has been used in existing experiments.

\section{Magnetoelastic coupling mechanism for acoustic and optical phonons}
\label{sect:coupling}

The displacement field of sound waves may be described by elastic  strains which couple to the CEF quadrupolar (or higher order multipolar) moments of the 4f shell. This causes a temperature dependent renormalisation of sound velocities (elastic constants) \cite{Thalmeierbook} and under suitable conditions a Faraday rotation of the polarisation \cite{Thalmeier78} due to the splitting of complex conjugate (R,L) acoustic modes. It is natural to expect a similar effect for doubly degenerate complex conjugate {\it optical} phonons. This has indeed been found in Ref.~\onlinecite{Schaack77} with Raman scattering in a magnetic field and explained in Ref.~\onlinecite{Thalmeier77}. An interesting possibility arises in non-Bravais lattices with an atomic basis: In the case that acoustic and optical phonons for \v q along an axis cause distortions of the same symmetry they will be coupled, both directly  and indirectly via quadrupole excitations of 4f CEF states. Such an a-o coupling will lead to an additional contribution to the Faraday rotation which has not been considered before. This new contribution turns out to solve the problem of the frequency dependence of the rotation angle, both for TGG and the previous case of CeAl$_2$.\\

Our starting Hamiltonian therefore has to include a term ($H_{ph}$) describing both acoustic and optical phonons. In addition it contains a part ($H_{4f}$) corresponding to the CEF- split 4f electrons and a phonon-4f electron interaction part $H_{ph-4f}$. The  coupled system is then described by
\bea
H=H_{ph}+H_{4f}+H_{ph-4f}
\eea
%
 
 
First we turn to the phonon part. Restricting to modes with \v q along $\hat{\v z}$ they may be written in terms of acoustic ($\phi_{\vq \mu}=a_{\vq \mu}+a^\dagger_{-\vq \mu}$) and optical  ($\Phi_{\vq \mu}=A_{\vq \mu}+A^\dagger_{-\vq \mu}$) phonon coordinates where $\mu = x,y$ denotes the two polarisations and 
$a^\dagger_{\vq \mu}, A^\dagger_{\vq \mu}$ are the usual phonon creation operators.
\bea
\epsilon_{\mu z}(i)=\sum_\vq Q_\vq^a\phi_{\vq \mu}e^{i\vq\vR_i};\qquad
Q_{\mu}(i)=\sum_\vq Q_\vq^o\Phi_{\vq \mu}e^{i\vq\vR_i}
\eea
Here $\vR_i$ denotes the lattice sites, furthermore we defined (N,M = number and mass of unit cell, respectively).
\bea
Q^a_{\v q}=iq(2MN\omega^a_{\v q})^{-\frac{1}{2}};\qquad
Q^o_{\v q}=(2MN\omega^o_{\v q})^{-\frac{1}{2}}
\label{eq:phonamp}
\eea
The total phonon Hamiltonian is then given by
\bea
H_{ph}=\sum_{\vq\mu}\omega^a_{\vq}[a^\dagger_{\vq \mu}a_{\vq \mu}+\frac{1}{2}]+
\sum_{\vq\mu}\omega^o_{\vq}[A^\dagger_{\vq \mu}A_{\vq \mu}+\frac{1}{2}] + 
\sum_{\vq\mu}\lambda_\vq\Phi_{\vq \mu}\phi^\dagger_{\vq \mu}
\label{eq:ph}
\eea
Here we defined $\lambda_\vq=iq(E/4M)(\omega^a_\vq\omega^o_\vq)^{-1/2}$ where E is the acoustic-optical (a-o) coupling constant between long wavelength acoustic strains and internal displacements of the optical mode. Note that in this representation $H_{ph}$ has not been completely diagonalised since there is an interaction term reflecting the fact that acoustic and optical phonons belong to the same type of representations of the group $G_\vq$ of the wave vector. 

The lattice vibration modes of the garnet structure RE$_3$Al$_5$O$_{12}$ which is isostructural to TGG (space group $O_h^{10}$) have been classified and discussed in Ref.~\onlinecite{Papagelis02}. Due to the large unit cell there are 240 vibrational modes for $\vq \rightarrow 0$. In addition to the acoustic T$_{1u}$ mode (one longitudinal, two transverse)  there are 17 T$_{1u}$ and 14 T$_{1g}$ optical modes and in addition 14 T$_{2g}$ and 16 T$_{2u}$ optical modes. For finite $\vq \parallel [001]$ the symmetry is reduced from cubic to uniaxial $C_{4v}$. Therefore the triply degenerate $T_1$ and $T_2$ modes split into singlet A and doublet E modes for both acoustic and optical phonons. The acoustic E modes correspond to the sound waves with (x,y) polarisation for propagation along z. Because of their E symmetry they will couple with {\it all} doubly degenerate optical E modes through the background force constants and the true acoustic eigenmodes are complicated superpositions of long wavelength and internal displacements. Therefore in a model based on magnetoelastic interactions  it is preferable to start with modes having pure acoustic strain and internal displacements and include the a-o coupling explicitly as in Eq.~(\ref{eq:ph}). Only in this case the magnetoelastic interaction can be written in terms of the simple strain and displacement amplitudes given in Eq.~(\ref{eq:phonamp}).
In the model we will include only one representative optical mode. This may be taken as the lowest $T_{1u}$ or $T_{2g}$ mode which is around $100$ $cm^{-1}$ in RE$_3$Al$_5$O$_{12}$ \cite{Papagelis02} and presumably also in TGG.

The 4f part of the Hamiltonian including the effect of an applied magnetic field is given by 
\bea
H_{4f}=\sum_{\alpha ni}E_\alpha|\Gamma_{\alpha n}\ra\la \Gamma_{\alpha n}|_i-h\sum_iJ_z(i) =
\sum_{\alpha n i}\tilde{E}_{\alpha n}|\tilde{\Gamma}_{\alpha_n}\ra\la\tilde{\Gamma}_{\alpha n}|_i
\eea
Where $E_\alpha$ and $|\Gamma_{\alpha n}\ra$ are the zero-field CEF level energies and states respectively and
$h=g_J\mu_BH (J = 6, g_J=\frac{3}{2}$) is the field variable.  The last expression in the above equation is written in terms of Zeeman split CEF energies  $\tilde{E}_{\alpha n}$ and states $|\tilde{\Gamma}_{\alpha n}\ra$. 

The lattice vibrations couple to the 4f states by changing the CEF potential due to induced local distortions \cite{Thalmeierbook}. In this way the E- type displacements of acoustic and optical modes will couple to E-type quadrupolar moments (O$_{xz}$,O$_{yz}$) of the 4f shell. Then the phonon-4f  or magnetoelastic Hamiltonian is given by
\bea
H_{ph-4f}=g_a\sum_{i\vq\mu}Q_\vq^a\phi_{\vq\mu}O_\mu(i)e^{i\vq\vR_i}+
g_o\sum_{i\vq\mu}Q_\vq^o\Phi_{\vq\mu}O_\mu(i)e^{i\vq\vR_i}
\label{eq:me}
\eea
where $\mu =x,y$ denotes polarisation for phonons and the quadrupolar operators O$_\mu$ ($\mu$=xz,yz,) are defined by
\bea
O_{xz}=J_xJ_z+J_zJ_x = \frac{1}{2}\bigl[(J_+J_z+J_zJ_+)+(J_-J_z+J_zJ_-)\bigr]        \no\\
O_{yz}=J_yJ_z+J_zJ_y = \frac{1}{2i}\bigl[(J_+J_z+J_zJ_+)-(J_-J_z+J_zJ_-)\bigr]
\label{eq:quadru}
\eea
with  $J_\pm=J_x\pm iJ_y$. Furthermore $g_a$ and $g_o$ are the magnetoelastic coupling constants for acoustic (a) and optical (o) modes respectively.

\section{Dyson equations for the phonon propagators}
\label{sect:dyson}

To obtain the propagating modes in the presence of magnetic ions and external field we have to set  up Dyson's equations for the phonon propagators of doubly degenerate a,o modes. As a first step, for better understanding of the a-o coupling term in Eq.~(\ref{eq:ph}) we  treat the case without 4f-phonon coupling and magnetic field. Then the phonon propagator is simply given by
\be
D^{-1}(\vq,\omega)=\left(\matrix
{\frac{1}{2\omega_\vq^a}(\omega^2-\omega^{a2}_\vq)  & -\lambda_\vq \cr
-\lambda^*_\vq & \frac{1}{2\omega_\vq^o}(\omega^2-\omega^{o2}_\vq)}
\right) 
\ee
where the diagonal elements are the unperturbed inverse phonon propagators for a,o modes. The nondiagonal terms mix the bare modes to the true vibrational modes whose frequencies are obtained from $det(D^{-1})=0$ or
\bea
(\omega^2-\omega^{a2}_\vq)(\omega^2-\omega^{o2}_\vq)=4\omega^a_\vq\omega_\vq^o|\lambda_\vq|^2
\eea
This leads to renormalised dispersions for both modes. Since we may assume $\omega^a_\vq\ll\omega^o_\vq$, at least in the long wave length limit the above equation leads to 
\bea
\tom_\vq^a=\omega_\vq^a\bigl[1-2\frac{|\lambda_\vq|^2}{\omega^{o2}_\vq}\frac{\omega^o_\vq}{\omega^a_\vq}\bigr];\qquad
\tom_\vq^o=\omega_\vq^o\bigl[1+2\frac{|\lambda_\vq|^2}{\omega^{o2}_\vq}\frac{\omega^a_\vq}{\omega^o_\vq}\bigr]
\eea
This shows the repulsion of a and o modes due to their interaction. In the long wavelength limit where $\omega_\vq^a=v_aq$ and $\omega^o_\vq=\omega_o-Dq^2$ we obtain a renormalised sound velocity for the acoustic mode 
\bea
\tilde{v}_a=v_a\bigl[1-\frac{2\tilde{\gamma}^2}{\omega_o}\bigr]=v_a\bigl[1-\frac{1}{2}\tilde{E}^2\bigr]
\label{eq:ao1}
\eea
where $\tilde{\gamma}^2\equiv (E/4M)^2/(v_a^2\omega_o)$ and $\tilde{E}^2=4\tilde{\gamma}^2/\omega_o$ is the dimensionless a-o coupling constant. The complementary correction to the optical phonon dispersion may be expressed as $\tom^o_\vq=\omega_o-\tilde{D}q^2$ where
\bea
\tilde{D}=D\bigl[1-\frac{2}{D}\Bigl(\frac{\tilde{\gamma}v_a}{\omega_o}\Bigr)^2\bigr]
\eea
Thus the a-o coupling of acoustic strains and internal  strains characterised by the constant E leads to a reduction of sound velocity and a reduced  dispersion of the optical mode.\\

Now we treat the full problem with magnetoelastic interactions included. We will show that the a-o coupling (E) plays an essential role in the acoustic Faraday rotation mechanism. For finite field and $g_a, g_o\neq 0$ the two polarisations have to be included explicitly for both modes. Then the total phonon propagator may be written as a $4\times 4$ matrix
\be
D^{-1}(\vq,\omega)=\left(\matrix
{D_a^{-1}  & -\Lambda_\vq \cr
-\Lambda^\dagger_\vq  & D_o^{-1}}
\right) 
\ee
where the diagonal $2\times 2$ blocks are given by the following expressions (the momentum label \vq~ has been omitted for simplicity)
\be
D_s^{-1}(\omega)=\frac{1}{2\omega_s}\left(\matrix
{\omega^2-\omega_s^2 -2\omega_sS_d^s(\omega)  & -2i\omega_sS_h^s(\omega)\cr
2i\omega_sS_h^s(\omega)& \omega^2-\omega_s^2 -2\omega_sS_d^s(\omega) }
\right) 
\ee
where $s=a,o$ is the mode index. Furthermore the non-diagonal block describing a-o coupling is given by
\be
\Lambda(\omega)=\left(\matrix{
\hat{\lambda} 1  & -S_h(\omega)\cr
S_h(\omega)&  \hat{\lambda} 1}
\right) 
\ee
with $\hat{\lambda}=i(|\lambda | +S_d)$. This matrix is anti-Hermitian with $\Lambda^\dagger=-\Lambda$.The diagonal ($S_d^s$) and nondiagonal ($S_h^s$) magnetoelastic self 
energies for a given mode (s = a,o) and $S_h, S_d$ for the a-o coupling are obtained as \cite{Thalmeier78,Thalmeier77}:
\bea
S_d^s(\omega)&=&\frac{1}{2}\omega_s\tilde{g}^2_s\la\la\hat{O}_{xz}\hat{O}_{xz}\ra\ra'_{\omega};\qquad
S_h^s(\omega)=\frac{1}{2}\omega_s\tilde{g}^2_s\la\la\hat{O}_{xz}\hat{O}_{yz}\ra\ra''_{\omega}\no\\
S_d(\omega)&=&\frac{1}{2}(\omega_a\omega_o)^\frac{1}{2}\tilde{g}^2\la\la\hat{O}_{xz}\hat{O}_{xz}\ra\ra'_{\omega};\qquad
S_h(\omega)=\frac{1}{2}(\omega_a\omega_o)^\frac{1}{2}\tilde{g}^2\la\la\hat{O}_{xz}\hat{O}_{yz}\ra\ra''_{\omega}
\eea
Here we defined $\tilde{g}=(\tilde{g}_a\tilde{g}_o)^\frac{1}{2}$ and $\hat{O}_{\alpha\beta}=O_{\alpha\beta}-\la O_{\alpha\beta}\ra$. The double brackets denote the dynamical susceptibility 
which are explicitly given in Sect.~\ref{sect:faraday}. Furthermore we used the definitions
\bea
N|g_aQ_\vq^a|^2=\frac{1}{2}v_aq\tilde{g}_a^2 \quad\mbox{or}\quad  \tilde{g}_a^2=\frac{g_a^2}{c_aV_c}\no\\
N|g_oQ_\vq^o|^2=\frac{1}{2}\omega_o\tilde{g}_o^2 \quad\mbox{or}\quad  \tilde{g}_o^2=\frac{g_o^2}{M\omega_o^2}\no\\
\eea
where $c_a=\rho v_a^2$ is the elastic constant (in the cubic garnets $c_a=c_{44}$) with $\rho = M/V_c$ being the mass density ($M,V_c$= unit cell mass and volume, respectively). 

The new phonon frequencies modified by magnetoelastic coupling as well as the a-o coupling are then obtained by finding solutions of $det D^{-1}(\vq,\omega)=0$. For that purpose it is convenient to transform the matrix propagator to circular polarised phonon  coordinates L(+) and R(-) according to $\phi_+=\frac{1}{\sqrt{2}}(\phi_x-i\phi_y)$ and  $\phi_-=\frac{i}{\sqrt{2}}(\phi_x+i\phi_y)$ for acoustic and similar for $\Phi_\pm$ circular optical modes. Then we can regroup the $4\times 4$ propagator in pairs of ($\phi_-,\Phi_-$) and  ($\phi_+,\Phi_+$) modes which leads to the decoupled $2\times 2$ circular mode propagators
\be
D_\pm^{-1}(\vq,\omega)=\left(\matrix
{\frac{1}{2\omega_{a\vq}}(\omega^2-\tilde{\omega}_{a\vq}^{\pm 2})  & -\Lambda^\pm_\vq \cr
-\Lambda^{\pm *}_\vq & \frac{1}{2\omega_{o\vq}}(\omega^2-\tilde{\omega}_{o\vq}^{\pm 2})}
\right) 
\label{eq:circprop}
\ee
where the renormalised circular mode frequencies are given by (s=a,o)
\bea
\tilde{\omega}^\pm_{s\vq}&=&\tilde{\omega}_{s\vq}^2\pm 2\omega_{s\vq} S_h^s(\omega)\no\\
\tilde{\omega}_{s\vq}^2&=&\omega_{s\vq}^2[1-\frac{2}{\omega_{s\vq}}S_d^s(\omega)]
\label{eq:split}
\eea
and the effective a-o coupling is obtained as
\bea
\Lambda^\pm_{\v q}&=&i[|\lambda_{\v q}|+S_d(\omega)\pm S_h(\omega)]
\label{eq:aoeff}
\eea
The last two terms in this equation are the dynamical correction caused by magnetoelastic interactions to the static background a-o coupling $\lambda_{\v q}$.
The total remaining a-o coupling in Eq.~(\ref{eq:circprop}) may be easily diagonalised by finding the zeroes of the determinant of the inverse $2\times 2$ propagators. This leads to the secular equations
\be
(\omega^2-\tilde{\omega}_{a\vq}^{\pm 2})(\omega^2-\tilde{\omega}_{o\vq}^{\pm 2})=
4\omega_{a\vq}\omega_{o\vq}|\Lambda^\pm_\vq|^2
\ee
For the  acoustic modes we may approximate $\omega_a\simeq \tilde{\omega}_{a\vq}\ll \tilde{\omega}_{o\vq}$. This leads to the renormalised circular mode frequencies
\be
\omega_a^2=\tilde{\omega}^2_{a\vq}\pm 2\omega_{a_\vq}S_h^a(\omega_{a\vq})
-4\omega_{a\vq}\omega_{o\vq}|\Lambda_\vq^\pm|^2\frac{1}{\tilde{\omega}_{o\vq}^{\pm 2}}
\label{eq:ARL}
\ee
The equivalent relations for the circular optical modes will not be considered further here. In the following we neglect the last term in Eq.~(\ref{eq:aoeff}) because for $\omega\rightarrow 0$ $S_h(\omega)\sim \omega$ and therefore it is negligible compared to $S_d(\omega)$.

\section{Faraday rotation angle}
\label{sect:faraday}

From the above result we may easily compute the Faraday rotation angle. In the Faraday configuration the frequency $\omega_a$ is fixed and L,R modes have different wavenumbers $q_L,q_R$ which may be derived from Eq.~(\ref{eq:ARL}) as
\bea
\omega_a^2&=&\tilde{v}_a^{-2}q_L^2-2v_aS_h^a(\omega_{a})q_L\no\\
\omega_a^2&=&\tilde{v}_a^{+2}q_R^2+2v_aS_h^a(\omega_{a})q_R
\label{eq:aLR}
\eea
where the renormalised sound velocities  of L,R acoustic modes are obtained in the limit $\omega_{a\vq}\rightarrow 0$ as
\bea
\tilde{v}_a^{\pm2}&=&\tilde{v}_a^2\bigl(1-\tilde{E}_t^2\frac{v_a^2}{\tilde{v}_a^2}\frac{\omega_o^2}{\tilde{\omega}_o^{\pm 2}}\bigr)\no\\
\tilde{v}_a&=&
=v_a[1-\tilde{g}_a^2\la\la\hat{O}_{xz}\hat{O}_{xz}\ra\ra'_0]\\
\tE_t&=&\tE+\tg\la\la\hat{O}_{xz}\hat{O}_{xz}\ra\ra'_0\no
\label{eq:renorm}
\eea
and the split optical mode frequencies $\tilde{\omega}_o^{\pm 2}$ are given by Eq.~(\ref{eq:split}). 
Here we defined the dimensionless bare a-o coupling constant by $\tE=(\tE_a\tE_o)^\frac{1}{2}$ with $\tE_a=E/(c_aV_c)$ and $\tE_o=E/(M\omega_o^2)$ which is equivalent to $\tE^2=4\tilde{\gamma}^2/\omega_o$ (see below Eq.~(\ref{eq:ao1})).We note again that $\tE_t$ contains the dynamical magnetoelastic corrections to the bare a-o coupling ($\tg=(\tg_a\tg_o)^\frac{1}{2}$). These corrections may be strongly field dependent as discussed in Sect.~\ref{sect:tgg}.
Under the approximation $\phi'=\frac{1}{2}(q_L-q_R)\ll \frac{1}{2}(q_L+q_R)\simeq q$ we obtain the acoustic Faraday rotation 
\be
\phi'=\frac{2v_aS^a_h}{\tv_a^{+2}+\tv_a^{-2}}+\frac{1}{2}\frac{\tv_a^{+2}-\tv_a^{-2}}{\tv_a^{+2}+\tv_a^{-2}}q
\ee
Using the explicit form of the quantities appearing in this expression  and $q=\frac{\omega_a}{v_a}=\frac{2\pi}{\lambda}$ where $\lambda$ denotes the ultrasonic wave length we get the final result 
\bea
\phi'=\phi'_a+\phi'_o=\frac{\pi}{\lambda}\Bigl(\frac{v_a}{\hv_a}\Bigr)^2
\Bigl[\tilde{g}_a^2\la\la\hat{O}_{xz}\hat{O}_{yz}\ra\ra''_{\omega_a}
+\tE_t^2\frac{1}{2}\bigl(\frac{1}{\hat{\omega}_o^{-2}}-\frac{1}{\hat{\omega}_o^{+2}}\bigr)\Bigr]
\label{eq:FA1}
\eea
with split optical phonon frequencies and renormalised sound velocities explicitly given by
\bea
\ho_o^{\pm 2}&=&1+\tg_o^2\la\la\hat{O}_{xz}\hat{O}_{xz}\ra\ra'_{\omega_o}
\pm\tg_o^2\la\la\hat{O}_{xz}\hat{O}_{yz}\ra\ra''_{\omega_o}\no\\
\hv_a^2&=&\tv_a^2\Bigl[1-\tE_t^2\bigl(\frac{v_a}{\tv_a}\bigr)^2\frac{1}{2}
\Bigl(\frac{1}{\ho_o^{-2}}+\frac{1}{\ho_o^{+2}}\Bigr)\Bigr]
\eea
The Faraday rotation in Eq.~(\ref{eq:FA1}) consists of two parts. The first one , $\phi'_a$ ($\sim\tg_a^2$) which was already discussed in Ref.~\onlinecite{Thalmeier78} is due to the direct coupling of long-wavelength strains to the quadrupole moments of the 4f shell. The second $\phi'_o$ ($\sim\tE_t^2$) and indirect part derived in the present work is due to the coupling of long-wavelength strains to the internal optical displacements which in turn couple to the 4f quadrupoles. The latter leads to a splitting of circular polarised optical phonons and subsequently (via $\tE_t$) to a splitting of circular acoustic modes. Therefore an indirect optical phonon contribution $\phi'_o$ to the acoustic Faraday rotation appears. We note that the common prefactor $\frac{\pi}{\lambda}\sim \omega_a$ in Eq.~(\ref{eq:FA1}) is linear in the sound frequency. Since the first term in the bracket is also $\sim \omega_a$ (see Eq.~(\ref{eq:qusus})) one has $\phi'_a\sim \omega_a^2$ for the direct acoustic contribution. This is because for $q\rightarrow 0$ the acoustic mode frequencies and {\it therefore also the splitting} of +,- (L,R) acoustic modes has to vanish. This contributes the additional frequency factor in $\phi'_a$. This is different for the indirect part $\phi'_o$. Because the splitting of optical $\hat{\omega}_o^\pm$ modes stays finite for $\vq \rightarrow 0$ the second term in the brackets of  Eq.~(\ref{eq:FA1}) is simply a (field dependent) constant and one finally gets  $\phi'_o \sim \omega_a$. For small $\omega_a$ the second part will always dominate and one has $\phi'\sim\omega_a$ for the total rotation angle. This solves the long-standing puzzle of the frequency dependence of acoustic Faraday rotation in paramagnetic 4f compounds.\\

In the following we give more explicit forms for the Faraday rotation by using the expression for the quadrupolar susceptibilities \cite{Thalmeier78,Thalmeier77}
\bea
\la\la\hat{O}_{\alpha}\hat{O}_{\beta}\ra\ra'_{\omega}&=&
{\sum_{lm}}'\frac{\tilde{\Delta}_{lm}(p_l-p_m)}{\omega^2-\tilde{\Delta}^2_{lm}}S_{\alpha\beta}^{lm}
+(kT)^{-1}\sum_lp_l\la l|\hat{O}_\alpha | l\ra\la l |\hat{O}_\beta | l\ra \delta_{\omega 0}\no\\
\la\la\hat{O}_{\alpha}\hat{O}_{\beta}\ra\ra''_{\omega}&=&
{\sum_{lm}}'\frac{\omega(p_l-p_m)}{\omega^2-\tilde{\Delta}^2_{lm}}A_{\alpha\beta}^{lm}
\label{eq:qusus}
\eea
Here $\tilde{\Delta}_{lm}=\tE_l-\tE_m$ are the excitation energies between CEF levels in an external field and $p_l =e^{-\tE_l/kT}/\sum_ne^{-\tE_n/kT}  $ are their occupation numbers. The prime denotes summation over terms with $\tE_l\neq\tE_m$. Furthermore $\hat{O}_\alpha=O_\alpha-\la O_\alpha\ra$ and the (anti-) symmetrised matrix elements are defined by 
\bea
S_{\alpha\beta}^{lm}&=&\frac{1}{2}[\la l |O_\alpha | m\ra\la m|O_\beta | l\ra + \la m |O_\alpha | l\ra\la l|O_\beta | m\ra]\no\\
A_{\alpha\beta}^{lm}&=&\frac{1}{2i}[\la l |O_\alpha | m\ra\la m|O_\beta | l\ra - \la m |O_\alpha | l\ra\la l|O_\beta | m\ra]
\label{eq:SA}
\eea
It is implied here that $|l\rangle,|m\rangle$ denote the eigenstates in the external field. In our case $\alpha,\beta = xz, yz$ and the notation will be simplified to $A_{xzyz}=A_{xy}$ and $S_{xzxz}=S_{xx}$ etc. Expanding the inverse of the split optical frequencies $\ho_o^{\pm 2}$ in the magnetoelastic coupling $\tg_o^2$ in Eq.(\ref{eq:FA1})
and inserting the expressions for quadrupolar susceptibilities we obtain after some algebra the Faraday rotation angle $\phi'=\phi'_a+\phi'_o$ as
\bea
\phi'=\frac{\pi}{\lambda}\Bigl(\frac{v_a}{\hv_a}\Bigr)^2
\Bigl[\tg_a^2{\sum_{lm}}'\frac{\omega_a(p_l-p_m)}{\omega_a^2-\tilde{\Delta}^2_{lm}}A_{xy}^{lm}+
\Bigl(\frac{\omega_o}{\tom_o}\Bigr)^4(\tE_t\tg_o)^2
{\sum_{lm}}'\frac{\omega_o(p_l-p_m)}{\omega_o^2-\tilde{\Delta}^2_{lm}}A_{xy}^{lm}\Bigl]
\label{eq:FA2}
\eea
Remembering that $\frac{\pi}{\lambda}\sim \omega_a$ this formula shows explicitly that the first part  $\phi'_a\sim \omega_a^2$ because we may take $\omega_a \rightarrow 0$ in the denominator: A typical sound frequency of $\nu_a=\omega_a/2\pi =0.5\mbox{GHz}$ corresponds to $\omega_a = 0.15$ K. This energy is much smaller than all other energy scales involved, in particular smaller than the lowest CEF excitation energy $\Delta_t =57.3$ K (see sect.~\ref{sect:tgg}) and even smaller than the thermal energy at the measuring temperature $1.4$ K which gives the scale of the CEF broadening. Therefore the above approximation is justified and the $\omega_a$ frequency dependence of the first part is simply given by $(\pi/\lambda)\omega_a\sim\omega_a^2$. In the second part due to indirect coupling to optical phonons the acoustic frequency appears only through the prefactor $(\pi/\lambda)$ leading to a linear behaviour $\phi'_o\sim \omega_a$ . One may estimate under which condition the linear (indirect optical) part dominates the quadratic (direct acoustic) part of the rotation angle. This happens when $\tg_a^2\omega_a \ll \Bigl(\frac{\omega_o}{\tom_o}\Bigr)^4(\tE_t\tg_o)^2 \omega_o$. Using $\tom_o\simeq\omega_o$ and assuming $\tg_a\simeq\tg_o, \tE_t\simeq\tE$ one obtains the condition $\tE\gg(\omega_a/\omega_o)^\frac{1}{2}$. For the typical $\omega_a/2\pi=0.5$ GHz and $\omega_o=150$ K this means we must have $\tE\gg 0.9\cdot 10^{-2}$. Assuming from Eq.~(\ref{eq:ao1}) that one has a modest change of sound velocity  $(v_a-\tv_a)/v_a\simeq 0.5\cdot 10^{-2}$ due to the bare a-o coupling only we get $\tE=0.1$, which is much bigger than the required value. Therefore we conclude that under any noticable a-o coupling the indirect ($\sim\omega_a$) contribution to the Faraday rotation will dominate the direct acoustic one ($\sim\omega_a^2$) and the sum ,i.e., the total observed Faraday rotation will be linear in $\omega_a$ to a good approximation.\\

The fact that optical phonon frequency $\tom_o$ and sound velocity $\hv_a$ are renormalised by the coupling to the diagonal quadrupolar susceptibilities enters only in a non-essential way through the modification of prefactors in Eq.~(\ref{eq:FA2}). Nevertheless we give their explicit expressions for completeness. The renormalised optical phonon frequency is obtained from the selfconsistent solution of
\bea
\tom_o^2=\omega_o^2
\Bigl[1-\tg_o^2{\sum_{lm}}'\frac{\tilde{\Delta}_{lm}(p_l-p_m)}{\tom_o^2-\tilde{\Delta}^2_{lm}}S_{xx}^{lm}\Bigr]
\label{eq:reo}
\eea
and the renormalised sound velocity from
\bea
\hv_a^2&=&v_a^2\Bigl[1-\tE_t^2
-\tg_a^2\Bigl({\sum_{lm}}'\frac{(p_m-p_l)}{\tilde{\Delta}_{lm}}S_{xx}^{lm}
+(kT)^{-1}\sum_lp_l\la l|\hat{O}_{xz} | l\ra\la l |\hat{O}_{xz} | l\ra \Bigr)
-(\tE\tg_o)^2{\sum_{lm}}'\frac{\tilde{\Delta}_{lm}(p_l-p_m)}{\tom_o^2-\tilde{\Delta}^2_{lm}}S_{xx}^{lm}\Bigr]
\label{eq:rea}
\eea
Note that the modified $\hv_a$  has three correction contributions: the first is due to the renormalised a-o coupling of elastic strains and internal displacements (see also Eq.~(\ref{eq:ao1})) while the second is due the direct magnetoelastic coupling and the third one due to indirect coupling to 4f quadrupoles via the optical modes. In the adiabatic limit when $\tom_o\ll\tilde{\Delta}_{lm}$ the third term has the same form as the second and just adds to an effective magnetoelastic coupling $\hat{g}_a^2=\tg_a^2+(\tE\tg_o)^2$. This was already noticed e.g. in Ref.~\onlinecite{Luethi79} where it was remarked that the contribution of the a-o coupling leads to an anomalously large  $\hat{g}_a^2$ as obtained from the temperature dependence of the sound velocity. It should be noted however that the adiabatic condition does not hold for TGG where the optical mode frequency is quite large as compared to the important CEF transition energies (Sect.~\ref{sect:tgg}). In this case the exact formula for $\hv_a(T)$ given in the above equation has to be used.

A remark on the solution of Eq.~(\ref{eq:reo}) is appropriate: In the nonresonant case where $\omega_o^2\neq\tilde{\Delta}_{lm}^2$ the renormalised $\tom_o$ and $\hv_a^2$ may be obtained by inserting the unperturbed $\omega_o$ on the right hand side of Eqs.~(\ref{eq:reo},\ref{eq:rea}). Due to the change of CEF levels with applied fields the resonance condition  $\omega_o^2\simeq\tilde{\Delta}_{lm}^2$  for some excitation $\tE_{l'} \rightarrow \tE_{m'}$ may be achieved at a special field $h_o$. In this case optical phonons and the CEF excitation will form 'mixed modes' which have no longer pure phononic or CEF excitation character.
Then Eq.~(\ref{eq:reo}) has to be solved selfconsistently. Neglecting the nonresonant contributions of all other excitations and defining $\tilde{\Delta}_{l'm'}^r=\Delta_{l'm'}(p_{m'}-p_{l'})S_{xx}^{l'm'} > 0$ one obtains:
\bea
\tom_o^2=\frac{1}{2}(\omega_o^2+\tilde{\Delta}_{l'm'}^2)+\sigma_{l'm'}
\bigl[\frac{1}{4}(\omega_o^2-\tilde{\Delta}_{l'm'}^2)^2+\tg_o^2\omega_o^2\tilde{\Delta}_{l'm'}^r\bigr]^\frac{1}{2}
\eea
where $\sigma_{l'm'}=sign(\omega_o^2-\tilde{\Delta}_{l'm'}^2)$. Passing through the anti-crossing point of CEF excitation $\tilde{\Delta}_{l'm'}(h)$ and phonon frequency $\omega_o$ the frequency of the phonon-type branch
changes only by a small finite amount given by $(\delta_o/\omega_o)_{T\rightarrow 0}^2=2(S_{xx}^{l'm'}\tg_o^2/\omega_o)^\frac{1}{2}\ll 1$ as long as $\tg_o^2/\omega_o\ll 1$.

\section{Application to TGG: CEF states and their Zeeman splitting}
\label{sect:tgg}
%
\begin{figure}
\includegraphics[width=8.0cm]{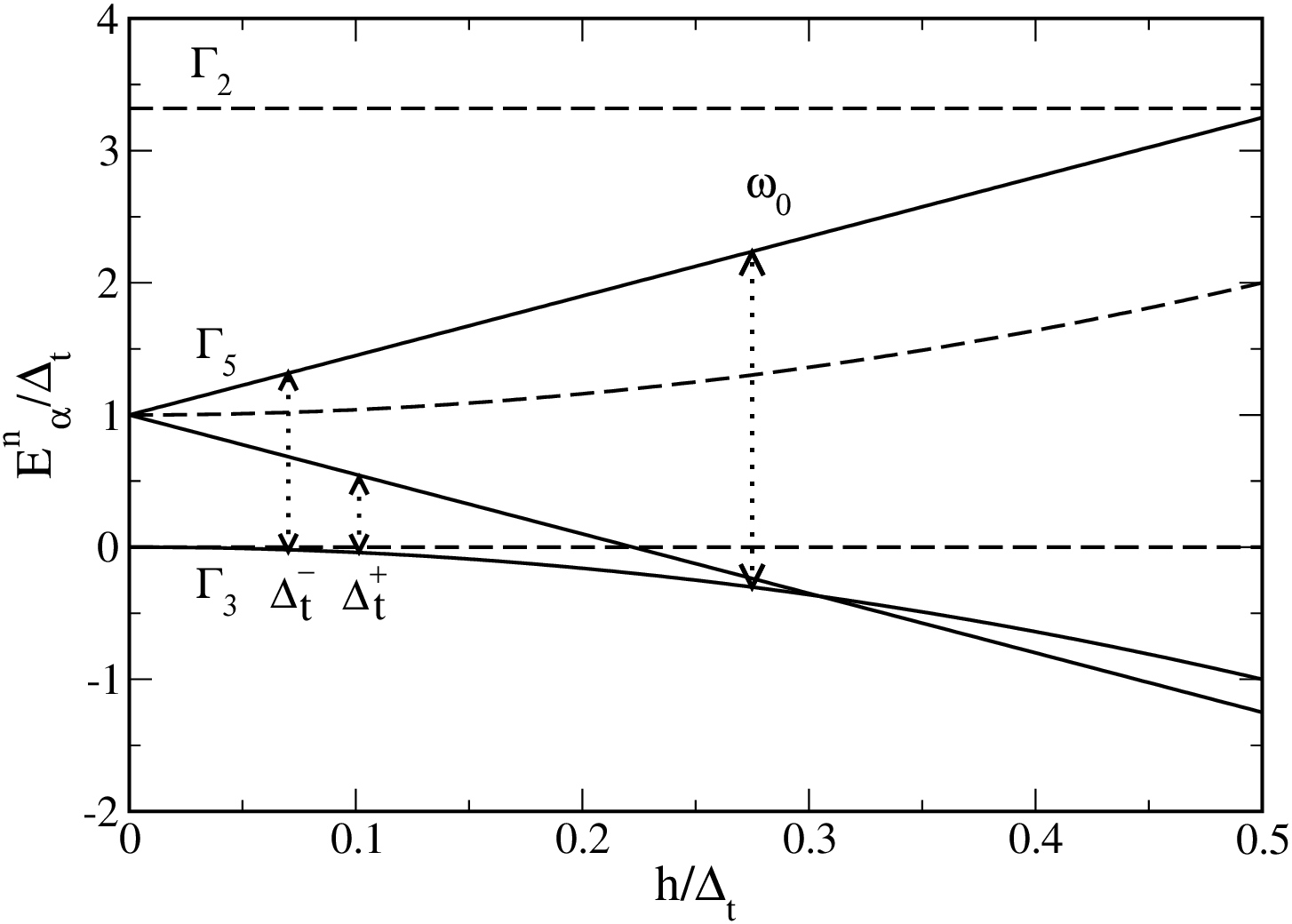}
\caption{Zeeman splitting of simplified cubic $\Gamma_3(0)-\Gamma_5(\Delta_t)-\Gamma_2(\Delta_s=3.3\Delta_t)$ 
level system. Only transitions between $\Gamma_3-\Gamma_5$ states corresponding to full lines have nonzero 
quadrupole matrix elements (Eqs.~(\ref{eq:Smat},\ref{eq:Amat})) and contributions to the quadrupole susceptibility. The empirical matrix elements used for the Zeeman splitting are $m_{55}^z=-4.5, m_{35}^z= -2.0, m^z_{25}=0.0$. Field scale: H = 28.6 T for
$h/\Delta_t$=0.5. The $\Gamma_3^1-\Gamma_5^1$ excitation ( $\Delta_t^-$) crosses the lowest optical phonon mode ($\omega_o=2.5\Delta_t$) around $h_o/\Delta_t \simeq 0.28$ ($\simeq$ 17 T) leading to a resonance in $\phi'_o$(h) (Fig.~\ref{fig:Fig2}).}
\label{fig:Fig1}
\end{figure}
%
A prerequisite to calculate the quadrupolar susceptibilities that enter the Faraday rotation angle is the knowledge of CEF states and their Zeeman splitting in a magnetic field. The O$_h^{10}$ space group symmetry of TGG leads to a D$_2$ site symmetry of Tb$^{3+}$ (J=6) non-Kramers ions. The corresponding CEF level scheme has been given in Ref.~\onlinecite{Guillot85}. It may be approximated by a cubic doublet-triplet-singlet level scheme $\Gamma_3$($\Delta_d$ = 0 K) - $\Gamma_5$($\Delta_t$ = 57.3 K) - $\Gamma_2$ ($\Delta_s$ = 190 K) \cite{Araki08} obtained for parameters W=-24.5 and x=0.8525 from the tables of Ref.~\onlinecite{Lea62}. Higher lying CEF states are neglected.

As far as eigenstates  $|\tilde{\Gamma}_\alpha^n\ra$ in a magnetic  field are concerned it is important to note that the linearly split triplet states $|\tilde{\Gamma}_5^{1,3}\ra =|\Gamma_5^{1,3}\ra$ and the  field-independent doublet component $|\tilde{\Gamma}_3^2\ra =|\Gamma_3^2\ra$ have the unperturbed wave functions (Fig.~\ref{fig:Fig1}) whereas the split-off ground state is mixed with a triplet component according to
\bea
|\tilde{\Gamma}_3^1=|\Gamma_3^1\ra+\frac{m_{35}^zh}{\Delta_t}|\Gamma_5^2\ra
\eea
The inclusion of the orthorhombic (D$_2$) CEF term for general field strength requires numerical calculations similar to Ref.~\onlinecite{Guillot85}. Now we need the matrix elements of the quadrupolar operators O$_{xz}$ and O$_{yz}$ of Eq.~(\ref{eq:quadru}) in the basis of Zeeman split CEF eigenstates $|\tilde{\Gamma}_\alpha^n\ra$. They only have to be considered in the subspace  
$S_3=\{|\tilde{\Gamma}_3^2\ra, |\Gamma_5^1\ra,|\Gamma_5^3\ra\} $ where they have non-zero matrix elements. The latter are given by
\bea
m_{11}&=&\la\Gamma_5^1|O_{xz}|\tilde{\Gamma}_3^1\ra = -i \la\tilde{\Gamma}_3^1|O_{yz}|\Gamma_5^1\ra\no\\
m_{13}&=&\la\Gamma_5^3|O_{xz}|\tilde{\Gamma}_3^1\ra =  i \la\tilde{\Gamma}_3^1|O_{yz}|\Gamma_5^3\ra
\eea
Using the (anti-) symmetrized matrices $A_{\alpha\beta}^{lm}, S_{\alpha\beta}^{lm}$ from Eqs.~(\ref{eq:Smat},\ref{eq:Amat}) in appendix \ref{sect:appA} this leads to explicit quadrupolar susceptibilities for the TGG CEF states:
\bea
\la\la\hat{O}_{xz}\hat{O}_{xz}\ra\ra'_{\omega}=\la\la\hat{O}_{yz}\hat{O}_{yz}\ra\ra'_{\omega}
&=&m_{11}^2R_d^-(\omega)(p_3^1-p_5^1)+m_{13}^2R_d^+(\omega)(p_3^1-p_5^3)\no\\
\la\la\hat{O}_{xz}\hat{O}_{yz}\ra\ra''_{\omega}=-\la\la\hat{O}_{yz}\hat{O}_{xz}\ra\ra''_{\omega}
&=&m_{11}^2R_h^-(\omega)(p_3^1-p_5^1)-m_{13}^2R_h^+(\omega)(p_3^1-p_5^3)
\label{eq:susTGG}
\eea
Note that for $T\rightarrow 0$ the CEF occupation differences $(p_l-p_m)$ approach unity. Furthermore we defined
\bea
R_d^\pm(\omega)&=&\frac{2\Delta_t^\pm}{\Delta_t^{\pm 2}-\omega^2};\qquad
R_h^\pm(\omega)=\frac{2\omega}{\Delta_t^{\pm 2}-\omega^2}
\label{eq:res} 
\eea
We have shown before that because of the different frequency dependences  $\phi'_a\ll\phi'_o$ for acoustic frequencies. Therefore 
$\phi'\simeq\phi'_o$ and we will discuss only the latter in the following. For $T\rightarrow 0$ we can write, using Eq.~(\ref{eq:qumat})
\bea
\label{eq:FA4}
\phi'_o&=&\frac{\pi}{\lambda}\Bigl(\frac{v_a}{\hv_a}\Bigr)^2F_o(\omega_o,H)\no\\
F_o(\omega_o,H)&=&\tilde{\kappa}_t^2\la\la\hat{O}_{xz}\hat{O}_{yz}\ra\ra''_{\omega_o}=
\tilde{\kappa}_t^2m_Q^2[\ham_-^2R_h^-(\omega_o,H)-\ham_+^2R_h^+(\omega_o,H)]\\
\tilde{\kappa}_t^2&=&\tilde{\kappa}_o^2\bigl(1+\rho_{ao}[\ham_-^2R_d^-(0,H)+\ham_+^2R_d^+(0,H)] \bigr)\no
\eea
with
$\tilde{\kappa}_o^2=\Bigl(\frac{\omega_o}{\tom_o}\Bigr)^4(\tE\tg_o)^2$ and $\rho_{ao}=\frac{2m_Q^2\tilde{g}^2}{\tE}$.  
In TGG we have $\Delta^-_t < \omega_o$ at zero field. However $\Delta_t^-$ increases linearly with field strength $(m^z_{55}<0)$ and will eventually cross the optical phonon frequency (Fig.~\ref{fig:Fig1}). Then the  Faraday rotation becomes large due to an optical phonon resonance in $R^-(\omega_o,H)$. Similarly  $\Delta^+_t$ decreases with increasing field and vanishes when the lowest triplet component crosses the ground state. In this case the effective a-o coupling strength diverges leading to an additional field dependence in the Faraday rotation caused by $R_d^+(0,H)$.
To avoid singular behaviour in the resonance case a finite linewidth of phonons ($\Gamma$) or CEF excitations ($\Gamma_t^\pm$) should be introduced. This can easily be done by replacing $R_{d,h}^\alpha(\omega,H)$ by the averaged quantities  $\tilde{R}_{d,h}^\alpha(\omega,H)$ which contains the finite linewidths (see appendix \ref{sect:appB}).
This leads to the final formula for the Faraday rotation in TGG which will be used for numerical calculation:
\bea
\phi'_o&=&\hat{\phi}'_o\hat{\kappa}_t^2(H)\Bigl[
\frac{2\omega_o^2\ham_-^2[\Delta_t^{-2}-\omega_o^2-\Gamma^2]}
{(\Delta_t^{-2}-\omega_o^2)^2+2\Gamma^2(\Delta_t^{-2}+\omega_o^2)+\Gamma^4}-
\frac{2\omega_o^2\ham_+^2[\Delta_t^{+2}-\omega_o^2-\Gamma^2]}
{(\Delta_t^{+2}-\omega_o^2)^2+2\Gamma^2(\Delta_t^{+2}+\omega_o^2)+\Gamma^4}\Bigr]
\label{eq:FA3}
\eea
Here the pre-factor $\hat{\kappa}_t^2=\tilde{\kappa}_t^2/\tilde{\kappa}_o^2$ describes the field dependence of the effective a-o coupling strength. It is determined by $\rho_{ao}=(2m_Q^2\tg^2/\tE)$ which gives a measure of the magnetoelastic corrections relative to the bare a-o coupling $\tE$. We obtain from the last of Eqs.~(\ref{eq:renorm}): 
\bea
\hat{\kappa}_t^2(H)&=&1+\rho_{ao}\Bigl(
\ham_-^2\frac{2\Delta_t^-}{\Delta_t^{-2}+\Gamma_t^{-2}}+
\ham_+^2\frac{2\Delta_t^+}{\Delta_t^{+2}+\Gamma_t^{+2}}\Bigr)
\label{eq:zetao}
\eea
If we ignore these corrections then $\hat{\kappa}_t^2(H)=1$ in Eq.~(\ref{eq:FA3}).
Furthermore the scale of the rotation angle is set through $\hat{\phi}'_o\sim\omega_a = 2v_a\frac{\pi}{\lambda}$ which is
explicitly given by
\be
\hat{\phi}'_o=\frac{\pi}{\lambda}\Bigl(\frac{v_a}{\hv_a}\Bigr)^2
\frac{(m_Q\tilde{\kappa}_o)^2}{\omega_o}
\label{eq:rotscale}
\ee
Note that the Faraday rotation is no longer singular for finite line widths $\Gamma$ and $\Gamma_t^\pm$. However it develops
a resonant behaviour as function of field which is determined by two effects: The optical phonon resonance with the singlet-triplet excitation at $\Delta^-_t(H)=\omega_o$ and the field dependence of the effective a-o coupling constant in Eq.~(\ref{eq:zetao}) when $\Delta^+_t(H)=0$. In TGG these conditions are fulfilled for approximately  the same field $H\simeq 17-20$ T (Fig.~\ref{fig:Fig1}) .
The above treatment may easily be generalised to include coupling to more than one optical phonon by summing over individual $n$-th phonon contributions like Eq.~(\ref{eq:FA3}) each characterised by a frequency $\omega_{o}^{(n)}$ and an effective a-o coupling strength $\tilde{\kappa}_{o}^{(n)}$ as well as $\rho^{(n)}_{ao}$.

%
\begin{figure}
\includegraphics[width=8cm]{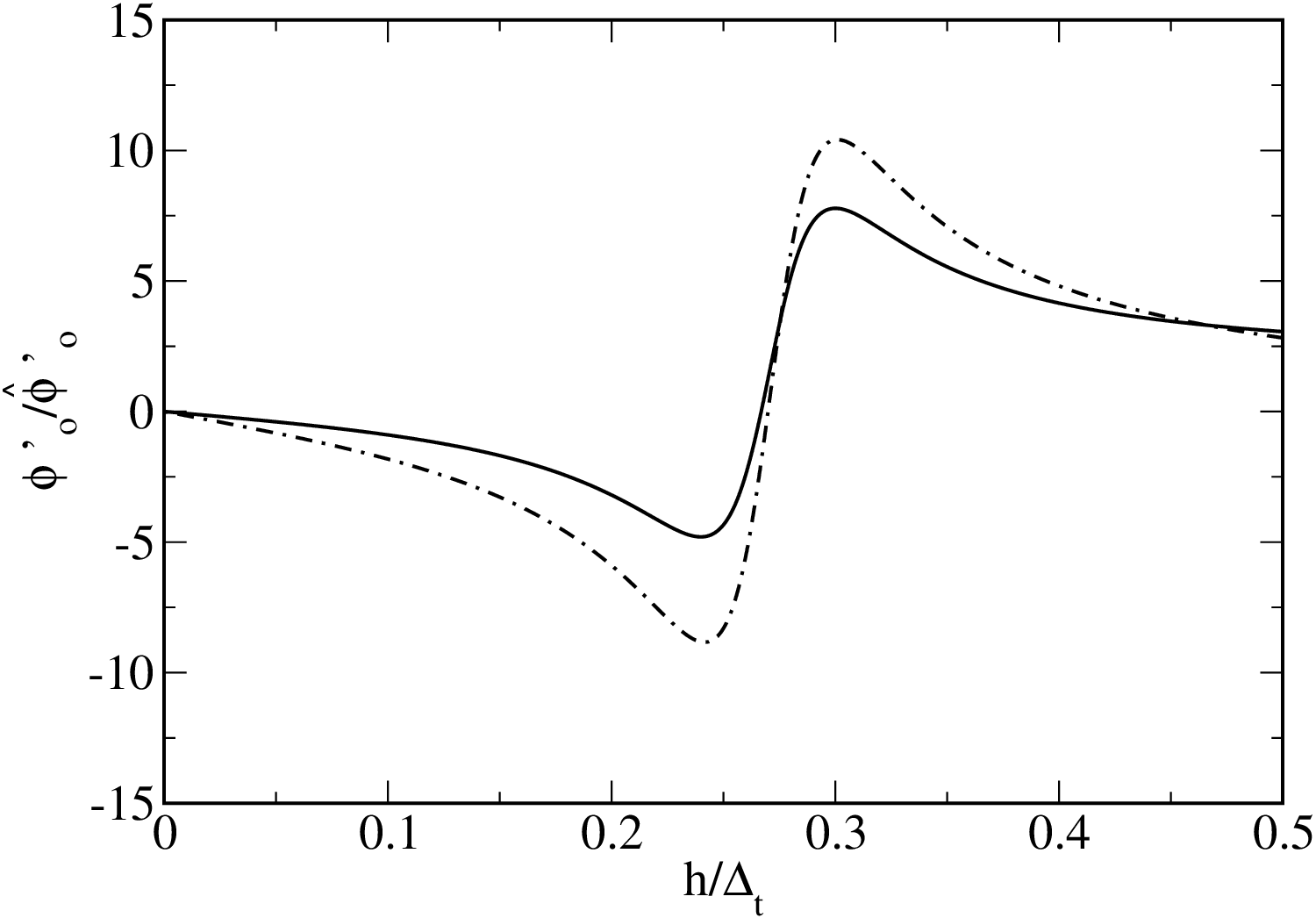}\hfill
\includegraphics[width=8cm]{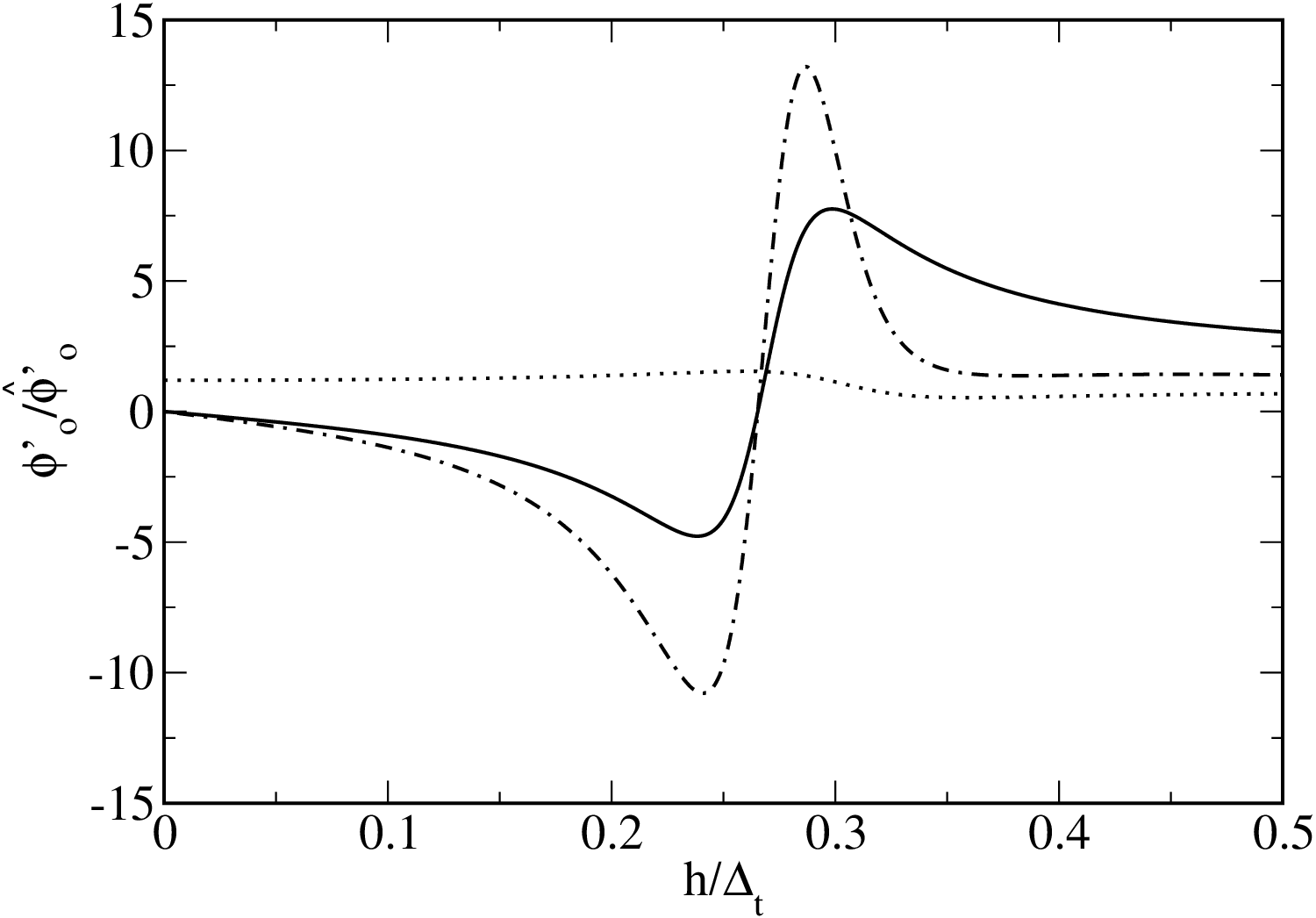}
\caption{Faraday rotation angle $\phi'_o/\hat{\phi}'_o$  as function of field strength ($\omega_o=2.5\Delta_t, \Gamma=0.2\Delta_t$).
Left (a): The optical phonon resonance appears around $\Delta_t^-= \omega_o$ (crossing with lowest optical phonon mode,
Fig.~\ref{fig:Fig1}). Full curve:  $m_Q'=0$. Dash-dotted line: $m_Q'=0.2$. The field dependence of quadrupolar matrix elements 
enhances the Faraday rotation. The bare a-o coupling ($\rho_{ao}=0$) is used in Eq.~(\ref{eq:zetao})
Right (b): Rotation angle for renormalised a-o coupling ($\rho_{ao}=0.05$, dash-dotted line) in comparison with bare case  ($\rho_{ao}=0$, full line). The enhancement factor $\hat{\kappa}_t^2(H)$ (dotted line) leads to an additional field dependence due to level crossing $\Delta_t^+ =0$ (Fig.~\ref{fig:Fig1}) which is close to the optical phonon resonance  $\Delta_t^- = \omega_o$. Here a CEF line width $\Gamma_t=0.1\Delta_t$ is used.}
\label{fig:Fig2}
\end{figure}
%

\section{Numerical results and discussion}
\label{sect:discussion}

Due to the low D$_2$ site symmetry the precise form of CEF states is not known with certainty. The cubic approximation proposed in Ref.~\cite{Araki08} using Ref.~\onlinecite{Lea62} is problematic because it leads to a linear Zeeman splitting of $\Gamma_5$ triplet which is far too small as compared to the calculation using the full D$_2$ CEF potential performed in Ref.~\cite{Guillot85}. For a calculation of the Faraday rotation angle according to Eq.~(\ref{eq:FA2}) within the latter approach it would be necessary to obtain the field dependent wave functions and energy levels numerically. To gain a principal insight a more simplified empirical approach was used here which allows to obtain closed expressions for the rotation angle in Eq.~(\ref{eq:FA3}). We kept the simplicity of the cubic level scheme and wave functions following Ref.~\onlinecite{Araki08}  and treated the dipolar and quadrupolar matrix elements as empirical parameters. The former may be fixed to obtain the qualitative level splitting and the latter enter only as scale factors that determine the strength of the resonance behaviour of $\phi'(h)$.\\

The simplified CEF level scheme is shown in Fig~\ref{fig:Fig1}. The essential states for the Faraday rotation are the split-off $\Gamma^1_3$ ground state component and the two linearly split $\Gamma_5$-states. Quadrupolar transitions between them lead to the rotation angle according to Eq.~(\ref{eq:FA3}). While the transition energy $\Delta_t^{+}$ decreases, $\Delta_t^{-}$  ($m^z_{55}<0$) increases with field strength. The field $h_{o}$ where the latter crosses the optical mode at $\omega_o=2.5\Delta_t$ is indicated by an arrow.\\ 

The Faraday rotation angle is shown in Fig.~\ref{fig:Fig2}. It vanishes for $h\rightarrow 0$ which is in contrast to the Faraday rotation for ferromagnets where the time reversal symmetry is already broken by the spontaneaous magnetisation. Under such condition the rotation angle is non-zero even at zero external field \cite{Guermeur67,Luethibook}.  At the field where $\Delta_t^-(h_o) \simeq \omega_o$ the rotation angle in Fig.~\ref{fig:Fig2}a  becomes resonant and changes sign when the field sweeps across. This is enhanced by the ground state level crossing when $\Delta_t^+=0$ at a slightly larger field which influences the effective renormalised a-o coupling (Fig.~\ref{fig:Fig2}b).\\

The qualitative  behaviour of Faraday rotation in Fig.~\ref{fig:Fig2} is very similar to the one observed in ferromagnets like Y$_3$Fe$_5$O$_{12}$ (YIG) \cite{Guermeur67}.  The shape of the resonance depends on the phonon and CEF excitation linewidths and its amplitude is enhanced by the opposite field dependence of the quadrupolar matrix elements $\ham_\pm$ (Eq.~(\ref{eq:qumat})) as well as a-o coupling renormalisation $\rho_{ao}$ (Fig.~\ref{fig:Fig2}).\\

\begin{figure}
\includegraphics[width=8.0cm]{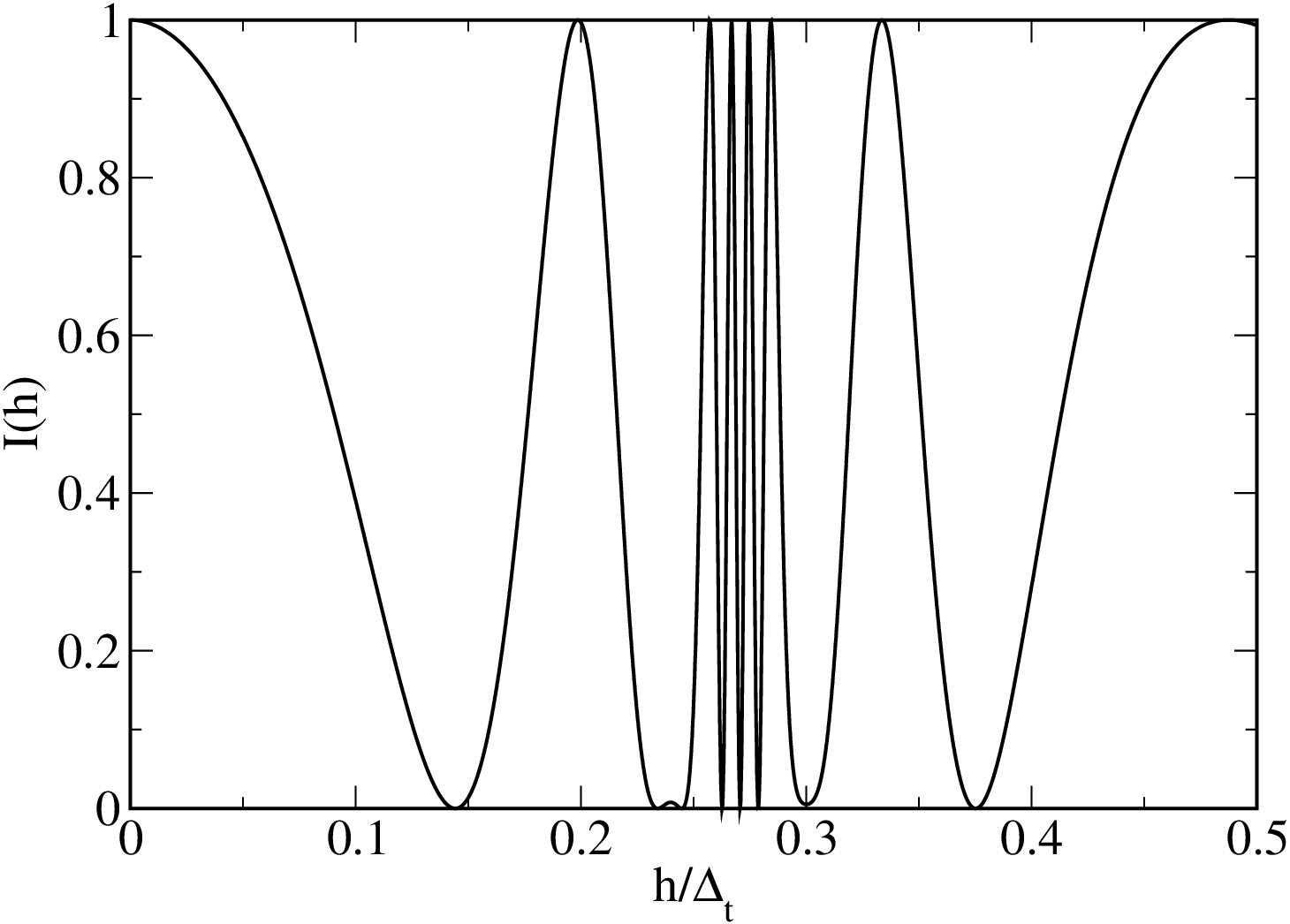}
\caption{Intensity of the soundwave component (Eq.~(\ref{eq:intens})) $\parallel \hat{x}$. The rapid oscillations occur 
for fields with large Faraday rotation angle $\phi'_o$ (Fig.~\ref{fig:Fig2}). Parameters are $\Gamma=0.2\Delta_t$ and $m'_Q =\rho_{ab}=0$ and $\hat{\phi}'_o = z_0 \equiv 1$ ($z_0$ = sample length).}
\label{fig:Fig3}
\end{figure}

In an experiment the rotation angle $\phi'(h)$ cannot be observed directly but is infered from the intensity modulation
\be
I(h;z_0)=A^2\cos^2[\phi'(h)z_0]
\label{eq:intens}
\ee
of the x-polarisation component of the propagating sound wave as it is detected at the end of the sample of length $z_0$. The typical intensity oscillations are shown in Fig.~\ref{fig:Fig3}. When the field sweeps through the resonance region the Faraday rotation becomes large and therefore the oscillations become rapid. This corresponds qualitatively to the experimental observations where the resonance is around $H=17-20$ T. In the latter the amplitude in the resonance region will also be strongly damped which is not described by the present treatment. Therefore instead of measuring the amplitude or intensity one may also measure the complementary damping which is large in the resonance region and also exhibits the oscillations due to Faraday rotation. If one has more optical phonons with different $\omega^{(n)}_o$ which couple to the acoustic modes the region of rapid Faraday rotation may exist in a more extended field region. The damping of the Faraday amplitude in the optical phonon resonance region can in principle be obtained by using the propagators with complex optical phonon poles at $\omega_o+i\Gamma$ and solving for complex wave numbers $q_L, q_R$.\\

Finally we mention that the present theory also explains the Faraday rotation in the case of paramagnetic CeAl$_2$ where the effect and its linear frequency dependence has first been observed.  In this compound a strong a-o coupling may be concluded from the large temperature effects which point to an enhanced effective magnetoelastic coupling due to optical phonons \cite{Luethi79}. Indeed in CeAl$_2$ the Ce atoms form a diamond type sublattice which has acoustic ($T_{1u}$) and lowest optical ($T_{2g}$) phonons with the same E ($C_{4v}$) symmetry for wave vector \vq =(q,0,0). This enables the a-o coupling as in TGG. In fact inelastic neutron scattering experiments \cite{Reichardt84} found that a-o coupling in this compound is strong enough to cause a large anticrossing effect of acoustic and optical phonons for q about halfway to the zone boundary. Therefore the Faraday rotation model discussed here is also  perfectly applicable for this compound. It should be mentioned however that the perturbative approach which was used here to solve Dyson's equations in an external field may not be adequate for CeAl$_2$ due to bound state formation of optical phonons and CEF excitations caused by large magnetoelastic coupling.\cite{Thalmeier82}

%
%

\section{Conclusion and Outlook}

The origin of acoustic Faraday rotation in paramagnetic 4f compounds has remained mysterious for a considerable time. Simple magnetoelastic theories consistently predicted that the rotation angle per unit length  should behave like $\phi' \sim \omega_a^2$ whereas experimental results \cite{Luethi79,Sytcheva10} invariably have shown that  $\phi' \sim \omega_a$ in the case where different sound frequencies $\omega_a$ have actually been used. 

In this work we have found the origin of this discrepancy and proposed a theoretical model for TGG which resolves the issue. In non-Bravais lattices acoustic and optical phonons may have the same symmetry with respect of the group of the wave vector \v q. In this case long wave length acoustic and internal optical displacements are intrinsically coupled both via background elasticity and the magnetoelastic coupling to CEF states. In such a case the acoustic Faraday rotation cannot be separated from the field splitting of optical phonons in a magnetic field. This leads to indirect contributions $\phi_o'$ to the former which are proportional to the effective a-o coupling squared. These contributions were shown to lead to a behaviour  $\phi_o'\sim\omega_a$ in agreement with experimental observation. They dominate the purely acoustic contribution  $\phi_a'\sim\omega_a^2$ due to the smallness of sound frequencies. Furthermore it was found that in TGG the crossing of CEF triplet levels with the lowest optical phonons and with the lowest doublet ground state component leads to resonant behaviour in the Faraday rotation angle for a field strength $H~\sim 17-20$ T. This has indeed been found in static and pulsed field experiments \cite{Sytcheva10}.\\ 

We believe the mechanism proposed here may be of more general validity. Certainly it is present in CeAl$_2$ where the paramagnetic acoustic Faraday rotation was first found. Indeed the diamond Ce sublattice of this compound ensures that acoustic and low-lying optical phonons have the same symmetry and are therefore strongly coupled for propagation along the cubic axis. Because Ce$^{3+}$ is a Kramers ion as opposed to Tb$^{3+}$ its CEF level scheme and therefore details of the theory are, however, quite different.

Finally we would like to propose that the field dependence of optical phonons in TGG which approximately extend from $90-900 cm^{-1}$ should be investigated by IR absorption and Raman scattering in high fields. According to our analysis there should be a field induced splitting of some of the doubly degenerate optical phonon modes which is indirectly responsible for the acoustic Faraday rotation. The observation of this splitting, in particular for the low frequency optical modes would be a direct support for the theory presented here.

\section*{Acknowledgements}
The author would like to thank A. Sytcheva and S. Zherlitsyn for many helpful discussions and for  making their experimental results available prior to publication. The author also thanks U. L\"ow and B. L\"uthi for helpful comments.

\appendix
\section{}
\label{sect:appA}
Here we derive the CEF energies and quadrupolar matrix elements of the field-split TGG (cubic) level scheme. Using the explicit form of CEF states given in the Lea, Leask and Wolf (LLW) tables \cite{Lea62} , the real matrix elements of the dipolar operator $J_z$ in the Zeeman term are given by 
\bea
m^z_{55}&=&\la\Gamma_5^1|J_z|\Gamma_5^1\ra = - \la\Gamma_5^3|J_z|\Gamma_5^3\ra \no\\
m^z_{35}&=&\la\Gamma_5^2|J_z|\Gamma_3^1\ra \\
m^z_{25}&=&\la\Gamma_5^2|J_z|\Gamma_2\ra \no
\eea
All other matrix elements vanish. The dipolar matrix elements calculated for the cubic CEF states are unreliable due to the admixture of other states by the D$_2$ part. Therefore we treat them as empirical parameters to reproduce the qualitative Zeeman splitting of triplet states. 
For  magnetic field applied along a cubic axis the Zeeman energy is  $H_Z=-hJ_z$. Up to second order in h the CEF level splitting and shifts of doublet, triplet and singlet are obtained as
\bea
\tilde{E}_3^1&=&{\tilde{\Delta}_d}= \Delta_d-\frac{(m_{35}^zh)^2}{\Delta_t}\no\\
\tilde{E}_3^2&=&\Delta_d=0\no\\
\tilde{E}_5^{1,3}&=&\Delta_t\mp m_{55}^zh \\
\tilde{E}_5^2&=&\tilde{\Delta_t} =\Delta_t-\frac{(m_{25}^zh)^2}{\Delta_s-\Delta_t} + \frac{(m_{35}^zh)^2}{\Delta_t}\no\\
\tilde{E}_2&=&\tilde{\Delta}_s =\Delta_s+\frac{(m_{25}^zh)^2}{\Delta_s-\Delta_t} \no
\eea
This leads to the symmetrized and antisymmetrized quadrupolar  matrices of Eq.~(\ref{eq:SA}) in the 
$S_3=\{|\tilde{\Gamma}_3^2\ra, |\Gamma_5^1\ra,|\Gamma_5^3\ra\} $  subspace according to
\be
S_{xz,xz}=S_{yz,yz}=\left(\matrix
{0  & m_{11}^2 & m_{13}^2 \cr
m_{11}^2& 0 & 0\cr
m_{13}^2& 0 & 0}\right);
\qquad A_{xz,xz}=A_{yz,yz} = 0
\label{eq:Smat}
\ee
\be
A_{xz,yz}=-A_{yz,xz}=\left(\matrix
{0  &- m_{11}^2 & m_{13}^2 \cr
m_{11}^2& 0 & 0\cr
-m_{13}^2& 0 & 0}\right);
\qquad S_{xz,yz}=S_{yz,xz} = 0
\label{eq:Amat}
\ee
The excitation energies  associated with non-zero matrix elements between ground state and two excited triplet components are given by (Fig.~\ref{fig:Fig1}):
\bea
\Delta_t^-&=&\tE_5^1-\tE_3^1=\Delta_t[1-\delta+\epsilon^2]\no\\
\Delta_t^+&=&\tE_5^3-\tE_3^1=\Delta_t[1+\delta+\epsilon^2]
\eea
where $\delta=\frac{m_{55}^zh}{\Delta_t}$ and  $\epsilon=\frac{m_{35}^zh}{\Delta_t}$. Furthermore the field dependent quadrupolar  matrix elements may be written as 
\bea
m_{11}&=&m_Q(1-\frac{m'_Q}{m_Q}\epsilon)\equiv m_Q\ham_-\no\\
m_{13}&=&-m_Q(1+\frac{m'_Q}{m_Q}\epsilon)\equiv -m_Q\ham_+
\label{eq:qumat}
\eea
where $m_Q$ is the zero-field singlet-triplet quadrupolar matrix element which may be expressed by the coefficients of the singlet-triplet wave functions. It enters into the proper scale Eq.~(\ref{eq:rotscale}) of the rotation angle.

\section{}
\label{sect:appB}
Finally we give the frequency dependences of diagonal and off-diagonal quadrupolar susceptibilities which have been averaged over the phonon and CEF spectral line shapes.
For the diagonal susceptibilities we introduce a linewidth $\Gamma_t^\pm$ for the relevant triplet excitations $\Delta_t^\pm$. Then we have ($\omega_a\rightarrow 0$)
\bea
\tR_d^\alpha(\omega_a,H)&=&\int_{-\infty}^{\infty} d\Delta\frac{\Gamma_t^\alpha/\pi}{(\Delta-\Delta_t^\alpha)^2+\Gamma_t^{\alpha 2}}
\frac{2\Delta}{\Delta^2-\omega_a^2}
=\frac{2\Delta_t^{\alpha}}
{(\Delta_t^{\alpha 2}+\Gamma_t^{\alpha 2})}
\eea
Likewise for the non-diagonal susceptibilities we introduce a linewidth $\Gamma$ for the optical phonon with frequency $\omega_o$ leading to ($\alpha=\pm$):
\bea
\tR^\alpha(\omega_o,H)&=&\int_{-\infty}^{\infty} d\omega\frac{\Gamma/\pi}{(\omega-\omega_o)^2+\Gamma^2}
\frac{2\omega}{\Delta_t^{\alpha 2}-\omega^2}
=\frac{2\omega_o[\Delta_t^{\alpha 2}-\omega_o^2-\Gamma^2]}
{(\Delta_t^{\alpha 2}-\omega_o^2)^2+2\Gamma^2(\Delta_t^{\alpha 2}+\omega_o^2)+\Gamma^4}
\eea
These expressions are used in Eq.~(\ref{eq:FA4}) to obtain the result in Eqs.~(\ref{eq:FA3},\ref{eq:zetao})


\begin{thebibliography}{99}

\bibitem{Matthews62}
H. Matthews and R.C. LeCraw
Phys. Rev. Lett. {\bf 8}, 397 (1962)

\bibitem{Guermeur67}
R. Guermeur, J. Joffrin, A. Levelut and J. Penn\'e,
Solid State Commun. {\bf 5}, 369 (1967)

\bibitem{Guermeur68}
R. Guermeur, J. Joffrin, A. Levelut, J. Penn\'e,
Solid State Commun. {\bf 6}, 519 (1968)

\bibitem{Kochelaev62}
B. I. Kochelaev,
Sov. Phys.  Solid State {\bf 4}, 1145 (1962)

\bibitem{Tucker80}
J. W . Tucker, 
J. Phys. C {\bf 13}, 1767 (1980)

\bibitem{Kittel58}
C. Kittel, 
Phys. Rev. {\bf 110}, 836 (1958)

\bibitem{Tucker73}
J. W. Tucker,
J. Phys. C {\bf 6}, 255 (1973)

\bibitem{Thalmeier78}
P.Thalmeier and P. Fulde,
Z. Phys. B {\bf 29}, 299 (1978)

\bibitem{Luethi79}
B. L\"uthi and C. Lingner
Z.Phys. B {\bf 34}, 157 (1979)

\bibitem{Sytcheva10}
A. Sytcheva, U. L\"ow, S. Yasin, J. Wosnitza, S. Zherlitsyn, T. Goto, P. Wyder and B. L\"uthi,
Phys. Rev. B (in press); arXiv:1006.0141

\bibitem{Thalmeierbook}
P. Thalmeier and   B. L\"uthi,
in "Handbook of the Physics and Chemistry of Rare Earths", vol. 14,
chap.69, p. 225, Elsevier, Amsterdam 1991

\bibitem{Schaack77}
G. Schaack,
Z. Phys. B {\bf 26}, 49 (1977)

\bibitem{Thalmeier77}
P. Thalmeier and P. Fulde,
Z. Phys. B {\bf 26}, 323 (1977)

\bibitem{Papagelis02}
K. Papagelis, G. Kanellis, S. Ves and G. A. Kourouklis,
phys. stat. sol. (b) {\bf 233}, 134 (2002)

\bibitem{Guillot85}
M. Guillot, A. Marchand, V. Nekvasil and F. Tcheou,
J. Phys. C: Solid State Phys. {\bf 18}, 3547 (1985)

\bibitem{Araki08}
K. Araki, T. Goto, Y. Nemoto, T. Yanagisawa and B. L\"uthi,
Eur. Phys. J. B {\bf 61}, 257 (2008)

\bibitem{Lea62}
K. R. Lea, M. J. M. Leask and W. P. Wolf,
J. Phys.Chem. Solids {\bf 23}, 1381 (1962)

\bibitem{Luethibook}
B. L\"uthi,
Physical Acoustics in the Solid State,
Springer, Berlin 2005

\bibitem{Reichardt84}
W. Reichardt and N. N\"ucker,
J. Phys. F {\bf 14}, L135 (1984)

\bibitem{Thalmeier82}
P. Thalmeier and P. Fulde
Phys. Rev. Lett. {\bf 49}, 1588 (1982)

\end{thebibliography}
\end{document}